\documentclass[11pt]{article}
\usepackage{amssymb}
\usepackage[fleqn]{amsmath}
\usepackage{setspace}
\usepackage[left=2.5cm,right=2.5cm,top=3cm]{geometry}

\newcommand{\cref}[1]{eqn.~(\ref{#1})}
\newcommand{\fn}{\footnote}

\begin{document}
\title{Can Galileons cause violation of the second law of black hole thermodynamics?}
\author{Saugata Chatterjee\fn{saugata.chatterjee@asu.edu}
\vspace{0.2in}\\
Arizona State University, \\
Tempe, Arizona 85281, USA \vspace{0.1in} \\
}

\date{}

\maketitle

\begin{abstract}
\noindent
Galileons are stable null energy condition (NEC) violating perturbations. If these NEC violating modes are coupled to Einstein gravity, they can cause violation of the second law of black hole thermodynamics. We demonstrate that galileons can only be coupled to quadratic gravity since they are generated by consistent Kaluza-Klein reduction of the Einstein-Hilbert-Gauss-Bonnet action from 10-dimensions to 4-dimensions. As a result, the NEC violation of galileons can no longer be the source of the violation of the second law of black hole thermodynamics.
\end{abstract}

\vspace{0.2in}

\noindent

\section{Introduction to Galileons}\label{sec:intro-galileons}
While attempting to explore the possibilities of modifying the long distance behavior of gravity by kinetically mixing a light scalar field, Nicolis et al. \cite{Nicolis:2008in} came across a very specific type of higher-derivative scalar field theory. This theory has the peculiar property that, in addition to the usual diffeomorphism degrees of freedom, it also has an additional degree of freedom, called the galilean symmetry. The galilean symmetry is reminiscent of the galilean transformation. This symmetry arose because the authors unmixed the scalar by a Weyl transformation. This means that they were dealing with a theory that is conformally coupled to gravity.  The conformal invariance of the theory manifests itself in the galilean symmetry of the scalar field and generates a series of higher-derivative scalar field  Lagrangians with the property that all of them have second-order equations of motion. The galileon (as obtained in \cite{Nicolis:2008in}) propagates on a flat background. Covariantizing the galileons turns out to be straightforward \cite{Deffayet:2009wt} but with the side-effect that it introduces ghosts in the theory. These ghosts can only be removed by adding certain non-minimal couplings of gravity to the galileon action, which in turn breaks the galilean symmetry explicitly. Therefore, nowadays galileon is a generic term for higher-derivative scalar field theories with second-order equations of motion.

In fact galileons are not the most general higher-derivative scalar field actions with second-order equations of motion. They are a subset of a much more general class of theories with higher-derivative, non-minimally coupled scalar field actions, with second-order equations of motion called the Horndeski theories \cite{Horndeski:1974wa}.
Let us demonstrate this connection by first writing down one of the Horndeski Lagrangians (eqn. 4.3 of \cite{Horndeski:1974wa}),
\begin{eqnarray}
 && L_2 = \sqrt{-g} \left( M_2(\phi,X)  R - 4  \dot M_2(\phi,X) \left(  (\Box \phi )^2 - \nabla_c \nabla_d \phi \nabla^c \nabla^d \phi  \right) \right) \label{eq:horndeski-L2}
\end{eqnarray}
where the $X =  (\partial \phi )^2 $  and the dot denotes differentiation with respect to $X$ i.e. $ \dot M_2(\phi,X) = \partial M_2(\phi,X) / \partial X$. $M_i(\phi ,X)$ are arbitrary functions of $\phi $ and $ (\partial \phi )^2 $. To demonstrate how the galileons end up being a special case of the Horndeski theories we write down the covariant galileons as obtained by \cite{Deffayet:2009wt},

\begin{eqnarray}
 L_4^{curved} = \sqrt{-g} (\partial \phi )^2 \left( 2 (\Box \phi )^2 - 2 \nabla_a \nabla_b \phi \nabla^a \nabla^b \phi - \frac{1}{2} (\partial \phi )^2 R \right) \label{eq:vikman-L4}
\end{eqnarray}

This Lagrangian contains a non-minimal coupling with the Ricci scalar, which is necessary to keep the equations of motion second-order when the background is curved. If we compare this Lagrangian with the Horndeski Lagrangian, \cref{eq:horndeski-L2}, it is clear that this Lagrangian is a special case of the Horndeski Lagrangian $L_2$ with the choice of the function $M_2 = -\frac{1}{2} X^2 = -\frac{1}{2}  (\partial \phi )^4 $.
Now we compare these Lagrangians with the original conformal galileons which were derived on a flat background \cite{Nicolis:2008in}.
\begin{eqnarray}
 L_4^{flat} =  \left( (\Box \phi )^2 (\partial \phi )^2  - (\partial \phi )^2 \nabla_c \nabla_d \phi \nabla^c \nabla^d \phi  + 2 \partial_a \phi  \nabla^a \nabla^b \phi \nabla_b \nabla^d \phi \partial_d \phi - 2 \Box \phi  \partial_c \phi \nabla^c \nabla^d \phi  \partial_d \phi  \right) \quad \label{eq:nicolis-L4}
\end{eqnarray}
The \cref{eq:nicolis-L4} is the flat space cousin of the Lagrangian \cref{eq:vikman-L4}. To see this  let us evaluate the following quantity,
\begin{eqnarray}
&&R_{ab} \partial^a \phi  \partial^b \phi (\partial \phi )^2 = R^{c}_{\phantom{c}acb}  \partial^a \phi  \partial^b \phi (\partial \phi )^2  = \left(  \partial^d \phi (\nabla^c \nabla_d \nabla_c \phi - \nabla_d \nabla^c \nabla_c \phi )  \right) (\partial \phi )^2 \nonumber \\
&&  = (\Box \phi )^2 (\partial \phi )^2  - (\partial \phi )^2 \nabla_c \nabla_d \phi \nabla^c \nabla^d \phi  - 2 \partial_a \phi  \nabla^a \nabla^b \phi \nabla_b \nabla^d \phi \partial_d \phi + 2 \Box \phi  \partial_c \phi \nabla^c \nabla^d \phi  \partial_d \phi  \qquad
\end{eqnarray}
where we have discarded some total derivatives at the end. It is clear that the two galileon Lagrangians are related as
\begin{eqnarray}
 L_4^{curved} = L_4^{flat} + G_{ab} \partial^a \phi  \partial^b \phi (\partial \phi )^2
\end{eqnarray}

This is the extra covariant coupling that needs to be added to the action in order to keep the equations of motion second-order. However the last term breaks the galilean invariance explicitly because the symmetry $\phi \rightarrow \phi  + b_a x^a + c $ is broken in the presence of the first derivatives of the scalar field. Looking at the Lagrangians (\cref{eq:horndeski-L2}) -- (\cref{eq:vikman-L4}) we notice one important thing. These Lagrangians contain at most second-order derivatives of the scalar field and no more. This is in fact a result of a theorem in the Horndeski paper \cite{Horndeski:1974wa} which states that the most general second-order equations of motion can be obtained from Lagrangians which are at most second-order in the derivatives of the fields. We also note the similarity with the Lovelock Lagrangians.
Even though the Lovelock theories are higher curvature theories of gravity, they still possess second-order equations of motion, and the action of these theories contain at most second-order in the derivatives of the metric. The connection is palpable, and we are led to conjecture that the galileon field might not be an arbitrary scalar field but one which is related to a particular degree of freedom of the metric. In fact this insight turns out to be true. The galileons can be obtained as the volume modulus of extra dimensions \cite{Charmousis:2008kc}  or as the position modulus of a brane embedded in a higher-dimensional bulk \cite{deRham:2010eu}.
We will call the previous one conformal galileon and the latter one DBI galileon (since the DBI action was used for the brane).
It has been shown in \cite{VanAcoleyen:2011mj} that the galileons can be obtained by Kaluza-Klein reduction of Lovelock theories over arbitrary cycles. This makes the galileons scalar field cousins of the Lovelock theories. They share the similar feature, that both of them are higher-derivative field theories without ghosts, which makes them part of a very attractive class of non-linear theories\fn{The effects of non-linearity introduced by the galileons can provide explanation for the dark forces at work at large scales \cite{DeFelice:2010pv,Tsujikawa:2010zza}.}.
In this paper, we will show how the galileons are recovered by the Kaluza-Klein reduction of the ten-dimensional Heterotic string theory\fn{They were first discovered in \cite{MuellerHoissen:1989yv} but the scalar terms were not recognized as galileons or as part of the Horndeski theories.}\cite{Charmousis:2008kc,Kobayashi:2004hq} (from five to four dimensions), and how this resolves a point of confusion in the literature regarding the connection between the null energy condition (NEC) and the stability of a non-linear theory.

\subsection{Connection between galileans, NEC violation, and stability}

The galileons possess NEC (null energy condition) violating solutions with a stable, Poincar\'{e} invariant vacuum. They are known to allow stable NEC violating perturbations to propagate at subluminal speeds. There is considerable interest in the literature regarding the NEC-stability connection and NEC-superluminality connection with the higher-derivative galileon theories being the test cases. Examples of theories which violate the NEC but have stable superluminal modes were found \cite{Dubovsky:2005xd}, and it was concluded that the correlation between NEC and stability is a weak one. However, this superluminality is consistent with unitarity \cite{deRham:2013hsa}. In fact the superluminal modes can be mapped onto subluminal modes when transforming from the Weyl to the DBI representation of the conformal galileons \cite{Creminelli:2013fxa}. It was soon realized  \cite{Creminelli:2006xe} that these NEC violating modes lead to a dispersion relation of the form  $\omega^2 = - a k^2 + b k^4$ which clearly has a gradient instability due to the wrong sign quadratic term coming from the kinetic part of the galileon field. However, the instability is avoided at short wavelengths (large wavenumber) due to the presence of the higher-derivative term in the action.
This severance of the connection between NEC and stability prompted others \cite{Creminelli:2012my} to explore  NEC $\leftrightarrow$ subluminality connection. The point was to find a NEC-violating background with a stable, Poincar\'{e} invariant vacuum so that the theory admits a NEC violating solution, which is stable under generic perturbations of the background, and which propagates subluminally. The trick is to parametrically distort the galileon genesis action \cite{Creminelli:2010ba} in a way that the NEC is violated. When the action is perturbed around a deSitter background, the gradient term has the right sign implying subluminality\fn{The sound speed is proportional to the coefficient of the gradient term: $c_s^2 = \frac{2- \alpha}{2 \alpha}$ where $\alpha$ is the coefficient of the $ (\partial \phi )^4$ term in the galileon action.}. Similar NEC violating, subluminal solutions were found in \cite{Hinterbichler:2012yn} for DBI galileons \cite{deRham:2010eu}.

If such stable NEC violating perturbations exist, and are coupled to the Einstein-Hilbert term only, then any black hole solution they source\fn{Black hole solutions with galileon hairs are obtained in \cite{Charmousis:2015txa}.} will no longer obey the second law of black hole thermodynamics. This is because, the galileons violating the NEC implies the violation of the Ricci convergence condition (by the Einstein's equation $R_{ab}k^a k^b = T_{ab}k^a k^b$), which in turn leads to the violation of the second law of black hole thermodynamics. We will demonstrate that the connection between the null energy condition and the Ricci convergence condition is severed because the galileons can be obtained (at sub-leading order in the $\alpha^\prime$ correction to the low energy effective action of the Heterotic string theory) by Kaluza-Klein reduction of the  Gauss-Bonnet term. Therefore, it is not justified to just couple the bare galileons to the Einstein-Hilbert term alone; the Gauss-Bonnet term must also be included. This severs the NEC argument because the NEC violation of the galileons can no longer source the violation of the Ricci convergence condition.

\section{Galileons from Kaluza-Klein reduction}\label{sec:galileons-kk}

The string effective action, in absence of loop corrections, is equal to the Einstein-Hilbert action. The equation of motion leads to  the equality $R_{ab} k^a k^b = T_{ab} k^a k^b $ and the null energy condition $T_{ab} k^a k^b \geq 0$ is implied by the Ricci convergence condition. However, when galileons are included, the null energy condition is violated, which means the Ricci convergence condition would be violated, and that can lead to violation of the second law of black hole thermodynamics. We will show below that there is no reason to assume that the galileans would be able to source stable NEC violating modes leading to violation of the Ricci convergence condition, and ultimately the laws thermodynamics (classically). This is because the galileans cannot be coupled to the Ricci alone; the Gauss-Bonnet invariant must also be present and it is this term which will prevent the stable, NEC violating modes, from violating the second law of black hole thermodynamics. In a nutshell, we are claiming that the galileons cannot be naively included into the tree level effective action, and the DBI genesis model \cite{Creminelli:2010ba} should be $\int (R + R_{GB} + galileons)$ rather than just $\int (R + galileons)$ and that the $\int (R + R_{GB} + galileons)$ action can be derived by consistent Kaluza-Klein reduction of the  Hetrotic string action to 4--dimensions.

In order to demonstrate our argument about the inclusion of the Gauss-Bonnet term, we begin by looking at the string effective action at sub-leading order in the $\alpha^\prime$ expansion \cite{Meissner:1996sa} (put in this form by appropriate field redefinitions)

\begin{eqnarray}
 \frac{1}{2 \kappa_{10}^2} \int d^{10}x \sqrt{-g_s} e^{-2 \phi} \left[ R + 4 (\partial \phi )^2 + \alpha^\prime \left( \frac{1}{8} (R_{abcd}^2 - 4 R_{ab}^2 + R^2)  - 2 G^{ab} \nabla_a \phi \nabla_b \phi + 2 \Box \phi (\partial \phi )^2 - 2 (\partial \phi)^4 \right) \right] \label{eq:heterotic}
\end{eqnarray}
This action is in string-frame. But if we look at the equations of motion, it is simple to convince ourselves that we can set the scalar field to zero consistently and the effective action reduces to (see appendix \ref{app:scalar-zero}),
\begin{eqnarray}
 \frac{1}{2 \kappa_{10}^2} \int d^{10}x \sqrt{-g_s} \left(  R  +  \frac{\alpha^\prime}{8} (R_{abcd}^2 - 4 R_{ab}^2 + R^2)  \right)  \label{eq:heteo-10D}
\end{eqnarray}

This is still in the string-frame as signified by the subscript ``s". Now we want to Kaluza-Klein \cite{Salam:1981xd,Hinterbichler:2013kwa} reduce this action. First, we have to reduce this ten--dimensional action to a five--dimensional action by compactifying the extra five dimensions over a five-torus, $T^5$. From there on, there are two ways to proceed. One way is to remain in five--dimensions, remove the non-minimal coupling from the Gauss-Bonnet term, and obtain galileons in five dimensions. The other way is to perform another Kaluza-Klein reduction over a $S^1$, and reduce this action to a four--dimensional action, and then remove the non-minimal coupling from the Einstein part, since the non-minimal coupling of curvature squared terms are non-removable in four--dimensions for dimensional reasons. The conformal transformation will still generate galileon terms but now there is a Gauss-Bonnet term non-minimally coupled to them.
Before proceeding it is interesting to note an interesting property specific to Lovelock terms which would simplify the analysis greatly. When Kaluza-Klein reducing the Lovelock terms (which includes the Einstein and the Gauss-Bonnet term) over an $S^1$, the kinetic terms of the scalar fields, which are generated in the action (while keeping the U(1) gauge fields fixed), turn out to be total derivatives.

\begin{eqnarray}
 && \frac{1}{2 \kappa_{D+1}^2} \int d^{D+1}x \sqrt{-g_{D+1}} \left(  R^{(D+1)}  +   \frac{\alpha^\prime}{8}  L_{GB}^{(D+1)}  \right) \nonumber \\
 &&  =  \frac{1}{2 \kappa_{D}^2} \int d^{D}x \sqrt{-g_{D}} ~ ~  e^{\phi } \left(  R^{(D)} - 2 ( (\partial \phi)^2 + \Box \phi )  +  \frac{\alpha^\prime}{8}  ( L_{GB}^{(D)} + 8 \tilde G^{ab} ( \nabla_a \nabla_b \phi + \partial_a \phi \partial_b \phi  ) ) \right) \nonumber \\
 &&  =  \frac{1}{2 \kappa_{D}^2} \int d^{D}x \sqrt{-g_{D}} ~ ~  e^{\phi } \left(  R^{(D)}  +  \frac{\alpha^\prime}{8} L_{GB}^{(D)} \right) \label{eq:Kaluza-Klein-D}
\end{eqnarray}

This action has only a non-minimal coupling of the scalar field. The equations of motion shows that this scalar field can also be set to zero consistently since there are no other matter fields present, and what is left is a lower dimensional action with the same functional form as the higher dimensional action. And this procedure can be repeated multiple times for every dimension and the ten-dimensional action can be reduced to a 5-dimensional action with 6 compact dimensions. The metric decomposition would look like this,
\begin{eqnarray}
  g_{\mu \nu} = \begin{pmatrix}
g^4_{ab} &           &         &         &  \\
        & exp(2\phi) &         &    0    &   \\
        &           &    1    &         &  \\
        &    0      &         & \ddots  &  \\
        &           &         &         & 1
\end{pmatrix}
\end{eqnarray}

\noindent 
We can now perform a Kaluza-Klein reduction of the 5--dimensional action over an $S^1$ with the volume $e^{\phi}$. The resulting action in four dimensions is the $D=4$ version of the last term in \cref{eq:Kaluza-Klein-D}.
\begin{eqnarray}
  \frac{1}{2 \kappa_{4}^2} \int d^{4}x \sqrt{-g_{4}} ~ ~  e^{\phi } \left(  R^{(4)}  +  \frac{\alpha^\prime}{8}  L_{GB}^{(4)} \right) \label{eq:Kaluza-Klein-5D}
\end{eqnarray}
A conformal transformation needs to be performed to remove the non-minimal coupling of the scalar field.
However, in four--dimensions the non-minimal coupling of the curvature squared terms cannot be removed by any conformal transformation. Hence the only option is to remove the non-minimal coupling of the Einstein term giving the action (appendix \ref{app:conf}).

\begin{eqnarray}
 \frac{1}{2 \kappa_{4}^2} \int d^{4}x \sqrt{-g_{4}} ~ ~ \left[   R^{(4)} - 2 (\partial \phi)^2  +  \frac{\alpha^\prime}{8} e^{-2 \phi} \left(  L_{GB}^{(4)}  +   16 G_{ab}^{(4)}  \partial^a \phi \partial^b \phi + 24 \Box \phi (\partial \phi )^2 \right)  \right] \label{eq:KK-4D}
\end{eqnarray}

This matches the one found by the authors in \cite{Kobayashi:2004hq}. 
However, this action does not contain all the terms from the galileon genesis action \cite{Creminelli:2010ba}. The $(\partial \phi )^4$ term is missing. The equations of motion are still second-order. This is because each term in \cref{eq:KK-4D} independently has a second-order equation of motion.

The equation of motion for this action can be easily obtained by using the generalized equation of motion as derived in the appendix \ref{app:gab-eom}. When we contract the equation of motion \cref{eq:eom} with null vectors we get,
\begin{eqnarray}
 R_{ab} k^a k^b + \frac{\alpha^\prime}{8}  e^{-2 \phi} H_{ab} k^a k^b = 2 (k \cdot \phi )^2 +  e^{-2 \phi} N_{ab} k^a k^b 
\end{eqnarray}
where $N_{ab} k^a k^b $ is the NEC violating part of the matter fields introduced due to the presence of galileons. However, the violation of the NEC by the galileon terms $N_{ab} k^a k^b$ does not imply the violation of the Ricci convergence condition, since they are no longer linearly related by the equation of motion.

\section{Discussion}
We have demonstrated that higher order, ghost free, kinetic terms of scalar fields, like the galileans when coupled with gravity, cannot appear with the Einstein term alone. The Gauss-Bonnet term is also present. Therefore, the NEC violation of the galileon terms does not imply the violation of the Ricci convergence condition, $R_{ab} k^a k^b \geq 0$, since the equations of motion are no longer $G_{ab} = T_{ab}$ but $G_{ab} + \alpha^\prime/8 H_{ab}= T_{ab}$, where $H_{ab}$ is the Gauss-Bonnet contribution to the equation of motion. Since the black hole entropy for higher curvature gravity is no longer simply proportional to the area of the event horizon, but proportional to a complex combination of curvature terms, the violation of the Ricci convergence condition is no longer related to the violation of the second law. The status of the second law of black hole thermodynamics is unclear. However, it is no longer clearly violated as before when the galileons were minimally coupled to the Einstein-Hilbert action.

In our analysis, we tried to simplify and reduce the proliferation of matter fields from the compactification process, by choosing Ricci flatness of the compact dimensions. While it is well known that dimensional reductions in string theory are carried out on K\"{a}hler manifolds, which are Ricci flat, it is interesting to note that in our case, it is to prevent the proliferation of fields, rather than getting the right kind of matter fields, that prompted the use of Ricci flat compact dimensions. It will be instructive to explore our compactification process without the simplifying assumption of Ricci flatness.

\section{Acknowledgment}
Author would like to thank Maulik Parikh for insightful discussions.

\appendix
\section{Conformal Transformation of the Ricci Scalar}\label{app:conf}
Given the conformal transformation of the metric $g_{ab}$ with a conformal factor $\Omega(x)$,
\begin{eqnarray}
 \tilde{g}_{ab} = \Omega(x)^2 g_{ab}
\end{eqnarray}
the Ricci scalar transforms as \cite{Wald:1984rg} (we will suppress the coordinate dependence of the conformal factor from now on),
\begin{eqnarray}
 \tilde{R} &= \Omega^{-2} \left( R - (D-2)(D-1) (\partial ln \Omega)^2 - 2 (D-2) \partial^2 ln \Omega \right) \nonumber \\
 R &=  \Omega^{2} \left( \tilde{R} - (D-2)(D-1) (\partial ln \Omega)^2 + 2 (D-2) \partial^2 ln \Omega \right)
\end{eqnarray}
The Einstein-Hilbert term which is the product $\sqrt{-g} R$, transforms as
\begin{eqnarray}
 & \sqrt{-\tilde{g}} ~ \Omega^{-D} \left(\Omega^{2}  \tilde{R} - (D-2)(D-1) (\partial \Omega)^2 + 2 (D-2)\Omega^{2}  \partial^2 ln \Omega \right) \nonumber \\
 & \quad =  \sqrt{-\tilde{g}}  ~\left(\Omega^{-(D-2)}  \tilde{R} + (- (D-2)(D-1) + 2 (D-2)^2 ) \Omega^{-D} (\partial \Omega)^2 \right.  \nonumber \\
 & \qquad \qquad \qquad \qquad \qquad \qquad \left.  + 2 (D-2)\partial( \Omega^{-(D-2)} \partial \Omega)  \right) \nonumber \\
 & \quad =  \sqrt{-\tilde{g}}  ~\left(\Omega^{-(D-2)}  \tilde{R} + (D-2)(D-3) \Omega^{-D} (\partial \Omega)^2 + \textrm{total derivative}  \right)
\end{eqnarray}

\section{Setting the scalar field to zero consistently}\label{app:scalar-zero}

The \cref{eq:heterotic} obtained in the string-frame is in the form
\begin{eqnarray}
\int d^{10}x \sqrt{-g_s}  \left[ e^{-2 \phi}  \left( R + g_1(\phi, \dot \phi, \ddot \phi ) \right) + e^{-2 \phi}  \left( L_{GB} + f_1(\phi, \dot \phi, \ddot \phi ) \right)  \right]
\end{eqnarray}
The equation of motion for the scalar field gets non-trivial curvature contributions from both the Einstein and the Gauss-Bonnet term. In $D>4$ dimensions, we can remove the scalar coupling $e^{-2 \phi} $ from the Gauss-Bonnet part using a conformal transformation to go to the Einstein frame (see appendix \ref{app:conf}).
\begin{eqnarray}
\int d^{10}x \sqrt{-g}  \left[ e^{\frac{2}{3} \phi} \left( R + g(\phi, \dot \phi, \ddot \phi ) \right) +  \left( L_{GB} + f(\phi, \dot \phi, \ddot \phi ) \right)  \right]
\end{eqnarray}
The equation of motion of the scalar field from this action has the form
\begin{eqnarray}
H(\phi, \dot \phi, \ddot \phi ) = \frac{2}{3}  e^{\frac{2}{3} \phi} R
\end{eqnarray}
where the first term contains combined derivatives of both the functions $g$ and $f$, which are  functions of only $\phi$ and its first and second derivative.  Now, if we assume that the background is Ricci flat, which is not an unreasonable assumption since many non-trivial solutions in gravity are Ricci flat, then the right hand side is identically zero, and we have a pure function of the scalar field which now admits a $\phi = 0$ solution. Therefore, we can consistently set the scalar field to zero in the action, provided we restrict ourselves to Ricci flat backgrounds.

\section{KK reduction of Einstein-Gauss-Bonnet term from D-dimensions to (D-1)-dimensions}\label{app:KK}
\subsection{KK reduction of Einstein-Hilbert term}
We start with a metric ansatz with the U(1) field set to zero.
\begin{eqnarray}
  g_{\mu \nu} = \left(    \begin{array}{c|c}
g_{ab} &   0         \\
\hline
0   & e^{2\phi }
\end{array} \right)
\end{eqnarray}
where $g_{yy} = e^{2\phi}$ is the component of the D--dimensional metric in the extra compact dimension. This looks like a natural foliation of the spacetime using a normal $\mathbf{n} = - N \mathbf{d} y $. We do not need the exact form of the normal or the normalization factor N which normalizes $n^a$ so that $\mathbf{n} \cdot \mathbf{n} = 1$. The only non-zero component of the 1-form is its y-component. This means that the metric is decomposed as
\begin{eqnarray}
 g_{ab} = h_{ab} + n_a n_b
\end{eqnarray} 
We write down the Gauss-Codazzi-Ricci equations.
\begin{eqnarray}
  {R^m_{\phantom{a}ijk}}^{(D)}  &=& {R^m_{\phantom{a}ijk}}^{(D-1)}  + (\mathbf n  \cdot \mathbf n )^{-1}  \left( K_{ij} K^m_k  - K_{ik} K^m_j \right) \\
  {R^n_{\phantom{a}ijk}}^{(D)}  &=& (\mathbf n  \cdot \mathbf n )^{-1} \left( K_{ik|j} -K_{ij|k}  \right) \\
  {R^n_{\phantom{a}ink}}^{(D)}  &=&  {\mathcal L}_n K_{ik} + K_{ip} K^p_k + {\mathcal D}_i A_k +  A_i A_k  \label{eq:riccimtw}
\end{eqnarray}
where $ {\mathcal L}_n $ is the Lie--derivative along $ n^a $ and ${\mathcal D}_i $ is the (D-1)--derivative. $ A_i  = n^k \nabla_k n_i $ is the acceleration of the (D-1)--hypersurface. The equations above are called Gauss, Codazzi, and Ricci equations respectively. Let us calculate some useful quantities.
\begin{eqnarray}
 A_i = n^\mu \nabla_\mu n_i = n^y \nabla_y n_i = n^y ( \partial_y n_i + \Gamma^y_{yi} n_y ) = (n^y n_y)  \Gamma^y_{yi} =  \Gamma^y_{yi} = \partial_i \phi
\end{eqnarray}
The extrinsic curvature $K_{ab}$
\begin{eqnarray}
 K_{ij}  = g_i^c g_j^d \nabla_c n_d = \nabla_i n_j
\end{eqnarray}
In the last line we expanded the projection operators $g_a^b$ in terms of the higher dimensional metric $g_\mu^\nu$ and the normal vectors. But the extra terms are all proportional to lower dimensional components of the normal vector, which are all zero. Expanding the above equation,
\begin{eqnarray}
 K_{ij}  = \nabla_i n_j  = \Gamma^y_{ij}  n_y = 0
\end{eqnarray}
since the Christoffel is zero. Therefore, all the extrinsic curvature terms and their contractions in the decomposition of the Riemann are zero. This leads to the following Gauss-Codazzi-Ricci equations.
\begin{eqnarray}
  {R^m_{\phantom{a}ijk}}^{(D)}  &=& {R^m_{\phantom{a}ijk}}^{(D-1)}  \label{eq:riemann-kk1}\\
  {R^y_{\phantom{a}ijk}}^{(D)}  &=& 0 \\
  {R^y_{\phantom{a}iyj}}^{(D)}  &=&  \nabla_i A_j + A_i A_j  = \nabla_i \nabla_j \phi + \partial_i \phi ~ \partial_j \phi
\end{eqnarray}
where all latin indices are lower dimensional ones and all Roman ones are higher dimensional. The Ricci tensor and the Ricci scalar are,
\begin{eqnarray}
 {R_{ij}}^{(D)} &=& {R^{m}_{\phantom{a}imj}}^{(D-1)}  + {R^{y}_{\phantom{a}iyj}}^{(D-1)}   =  {R_{ij}}^{(D-1)} + \nabla_i \nabla_j \phi + \partial_i \phi ~ \partial_j \phi  \\
 {R^y_{\phantom{a}y}}^{(D)}  &=& \Box \phi + (\partial \phi )^2 \\
 {R}^{(D)} &=& g^{ij} {R_{ij}}^{(D)} + g^{yy} {R_{yy}}^{(D)}  =  {R}^{(D-1)} + 2 \left(  \Box \phi + (\partial \phi )^2 \right) \label{eq:riemann-kk6}
\end{eqnarray}
Since none of the quantities depend on the extra dimension we can integrate over the extra dimension.
\begin{eqnarray}
 &&\frac{1}{2 \kappa^2_D} \int dy d^{D-1} x \sqrt{-g_D} R^{(D)} \nonumber \\
 &&\frac{1}{2 \kappa^2_{D-1}} \int d^{D-1} x \sqrt{-g_{D-1}} e^\phi  \left(  R^{(D-1)} + 2 \left( \Box \phi + (\partial \phi )^2 \right) \right)\nonumber \\
 &&\frac{1}{2 \kappa^2_{D-1}} \int d^{D-1} x \sqrt{-g_{D-1}} e^\phi   R^{(D-1)}
\end{eqnarray}
the scalar kinetic terms form a total derivative and so they are discarded.

\subsection{KK reduction of Gauss-Bonnet term}
Using \cref{eq:riemann-kk1} -- \cref{eq:riemann-kk6}, we can Kaluza-Klein reduce the Gauss-Bonnet term,
\begin{eqnarray}
 &&\frac{1}{2 \kappa^2_D} \int dy d^{D-1} x \sqrt{-g_D} {L_{GB}}^{(D)} \nonumber \\
 &&\frac{1}{2 \kappa^2_{D-1}} \int d^{D-1} x \sqrt{-g_{D-1}} e^\phi  \left( {L_{GB}}^{(D-1)} + 8 G^{ab} \left( \nabla_a \nabla_b    \phi + \partial_a \phi \partial_b \phi \right) \right)   \nonumber \\
 &&\frac{1}{2 \kappa^2_{D-1}} \int d^{D-1} x \sqrt{-g_{D-1}} e^\phi  {L_{GB}}^{(D-1)}
\end{eqnarray}
Again the last two terms form a total derivative as a result of the Bianchi identity. The Einstein tensor is constructed out of (D-1)--dimensional metric and the derivatives are also (D-1)--dimensional (See \cite{MuellerHoissen:1989yv}).

\section{Derivation of the equations of motion for 4--dimensional KK-reduced action}\label{app:gab-eom}

In this section, we derive the equation of motion for the following equation
\begin{eqnarray}
 \frac{1}{2 \kappa_{4}^2} \int d^{4}x \sqrt{-g_{4}} ~ ~ \left[   R^{(4)} - 2 (\partial \phi)^2  +  \frac{\alpha^\prime}{8} e^{\mu \phi} \left(  L_{GB}^{(4)} + 16 G_{ab}^{(4)}  \partial^a \phi \partial^b \phi + 24 \Box \phi (\partial \phi )^2 \right)  \right]
\end{eqnarray}
where we have replaced the non-minimal coupling $e^{-2 \phi}$ of \cref{eq:KK-4D} by $e^{\mu \phi}$, intending to derive the equation of motion for the most general non-minimal coupling and then replacing $\mu = -2$ at the end.

A straightforward variation of the non-coupled terms gives,

\begin{eqnarray}
\delta \int \left( R - 2 (\partial \phi)^2 \right) \Rightarrow G_{ab} - 2 \left( \partial_a \phi \partial_b \phi - \frac{1}{2} g_{ab} (\partial \phi)^2 \right) \delta g^{ab}
\end{eqnarray}

Similarly, the equation of motion for the last scalar term would be,
\begin{eqnarray}
\delta \int \left( e^{\mu \phi} \Box \phi (\partial \phi)^2  \right) \Rightarrow  e^{\mu \phi} \left( \nabla_a \nabla_b \phi  (\partial \phi)^2 + 2 \Box \phi  \partial_a \phi \partial_b \phi - \frac{1}{2}  g_{ab} \Box \phi (\partial \phi)^2 \right) \delta g^{ab}
\end{eqnarray}

The second and the third terms are found in \cref{app:gab-eom1} and \cref{app:gab-eom2}. Putting them together we get,
\begin{eqnarray}
&& G_{ab} - 2 \left( \partial_a \phi \partial_b \phi - \frac{1}{2} g_{ab} (\partial \phi)^2 \right)  +  \frac{\alpha^\prime}{8} e^{-2 \phi} \left(  \left( H_{ab} + Q_{ab} \right) - \frac{1}{4} g_{ab}  \left( L_{GB} + Q \right) \right) \nonumber  \\
&&  + 2\alpha^\prime e^{-2 \phi} \left( M_{ab} -  g_{ab} M \right) + 3\alpha^\prime e^{-2 \phi} \left( \nabla_a \nabla_b \phi  (\partial \phi)^2 + 2 \Box \phi  \partial_a \phi \partial_b \phi - \frac{1}{2}  g_{ab} \Box \phi (\partial \phi)^2 \right) = 0 \label{eq:eom}
\end{eqnarray}
Contracting the equation of motion with null vectors $k^a$ we get,
\begin{eqnarray}
 R_{ab} k^a k^b + \frac{\alpha^\prime}{8}  e^{-2 \phi} H_{ab} k^a k^b = 2 (k \cdot \phi )^2 +  e^{-2 \phi} N_{ab} k^a k^b 
\end{eqnarray}
where $N_{ab} k^a k^b$ is the galileon portion and $H_{ab} k^a k^b $ is the gravity portion of the equation of motion.
\begin{eqnarray}
N_{ab}  k^a k^b = \left( \frac{\alpha^\prime}{8} Q_{ab}  + 2\alpha^\prime M_{ab} \right)  k^a k^b   + 6\alpha^\prime  \Box \phi  (k \cdot \phi )^2  + 3\alpha^\prime k^a k^b  \nabla_a \nabla_b \phi  (\partial \phi)^2
\end{eqnarray}

\subsection{Derivation of the equations of motion for non-minimally coupled Gauss-Bonnet term}\label{app:gab-eom1}
The non-minimally coupled Gauss-Bonnet term can be derived without a brute force variation using the general formula in \cite{Chatterjee:2014qfa}.

\begin{eqnarray}
\delta \int \left( e^{\mu \phi}  L_{GB}^{(4)} \right) \Rightarrow  e^{\mu \phi} \left(  \left( H_{ab} + Q_{ab} \right) - \frac{1}{4} g_{ab}  \left( L_{GB} + Q \right) \right) \delta g^{ab}
\end{eqnarray}
where $H_{ab} - \frac{1}{4} g_{ab} L_{GB} $ is the equation obtained from the variation of the uncoupled Gauss-Bonnet term,
\begin{eqnarray}
H_{ab} &=& R R_{ab} - 2 R_{ac} R^c_b - 2 R^{cd} R_{acbd} + R_a^{cde} R_{bcde} \\
L_{GB} &=& \left(  R_{abcd}R^{abcd} - 4 R_{ab} R^{ab} + R^2 \right) \\
Q_{ab} &=&  \mu \left( 2  R_{d(a} \nabla_{b)} \nabla^d \phi + R_{acbd} \nabla^c \nabla^d \phi - \frac{1}{2} R \nabla_a \nabla_b \phi -  R_{ab} \Box \phi - \frac{\mu}{2} R \partial_a \phi \partial_b \phi \right. \nonumber  \\
&&  \left. + 2 \mu R_{d(a} \partial_{b)} \phi \partial^d \phi + \mu R_{acbd} \partial^c \phi \partial^d \phi - \mu R_{ab} (\partial \phi)^2 \right) \\
Q  &=&   \mu \left( -R_{cd} \nabla^c \nabla^d \phi + \frac{1}{2} R  \Box \phi + \frac{\mu}{2} R (\partial \phi)^2 - \mu  R_{cd}  \partial^c \phi \partial^d \phi  \right)
\end{eqnarray}
We have used a symmetrization factor: $A_{(bc)} = \frac{1}{2} (A_{bc} + A_{cb}) $. Note that when there is no non-minimal coupling, i.e. $\mu = 0$ , $Q_{ab}$ and $Q$ drop out of the equation of motion.

\subsection{Derivation of the equations of motion for  $e^{\mu \phi}G_{ab} \partial^a \phi \partial^b \phi $}\label{app:gab-eom2}

Doing a straightforward variation of just $G_{ab} \partial^a \phi \partial^b $ (without the square root of the metric) leads to,
\begin{eqnarray}
 \delta R_{ab}  \partial^a \phi \partial^b \phi + 2 R_{c(a} \partial_{b)} \phi  \partial^c \phi \delta g^{ab}  - \frac{1}{2} R \partial_a \phi \partial_b \phi \delta g^{ab}  - \frac{1}{2} (\partial \phi)^2 \delta R_{ab} g^{ab} - \frac{1}{2} (\partial \phi)^2 R_{ab}  \delta g^{ab}  \nonumber \\
 \label{eq:brute-var}
\end{eqnarray}
Adding the non-minimal coupling $e^{\mu \phi}$ and discarding the total derivatives while assuming the metric is continuous and differentiable at the boundaries, we arrive at the equation of motion. The derivation of the full equation of motion for $e^{\mu \phi} G_{ab} \partial^a \phi \partial^b \phi$ is tedious. We would just quote the result. 
\begin{eqnarray}
\delta \int \left( e^{\mu \phi} G_{ab} \partial^a \phi \partial^b \phi \right) \Rightarrow  e^{\mu \phi} \left( M_{ab} -  g_{ab} M \right) \delta g^{ab}
\end{eqnarray}
where we have defined,

\begin{eqnarray}
M_{ab} &=& \frac{1}{2}  \left( 4 R_{d(a} \partial_{b)} \phi \partial^d \phi - R \partial_a \phi \partial_b \phi - R_{ab} (\partial \phi)^2  -2 R_{acbd} \partial^c \phi \partial^d \phi + \mu \Box \phi \partial_a \phi \partial_b \phi \nonumber \right. \\
&& \left. + ( \mu (\partial \phi)^2 + 2 \Box \phi ) \nabla_a \nabla_b \phi  - 2 \mu \partial_d \phi (\nabla^d \nabla_a \phi ) \partial_b \phi - 2 \nabla_a \nabla^c \phi \nabla_b \nabla_c \phi \right) \\
M &=& \frac{1}{2}  \left(  \mu \Box \phi (\partial \phi)^2 - \mu \partial_c \phi (\nabla^c \nabla^d \phi ) \partial_d \phi  + \nabla^c \nabla^d \phi \nabla_c \nabla_d \phi + R_{cd} \partial^c \phi \partial^d \phi +  G_{cd} \partial^c \phi \partial^d \phi \right)
\end{eqnarray}


\providecommand{\href}[2]{#2}\begingroup\raggedright\endgroup

\end{document}